\documentclass{article}
\usepackage{geometry}                
\usepackage{graphicx}
\usepackage{stmaryrd}
\usepackage{amssymb}
\usepackage{amsthm, bm}
\usepackage{bbm}
\usepackage{amsmath, cancel, centernot }
\usepackage[mathscr]{eucal}
\usepackage{mathtools}
\usepackage{epstopdf}
\usepackage{mathrsfs}
\usepackage{newtxmath}

\title{A Deterministic Model of Free Will}
\author{Tim Palmer\\ Department of Physics, University of Oxford, UK\\
tim.palmer@physics.ox.ac.uk}
\date{\today}                                          
\makeatletter
\newcommand\be{\@ifstar{\[}{\begin{equation}}}
\newcommand\ee{\@ifstar{\]}{\end{equation}}}
\newcommand\bp{\begin{pmatrix}}
\newcommand\ep{\end{pmatrix}}

\newtheorem*{theorem*}{Theorem}

\makeatother
\begin{document}
\bibliographystyle{plain}
\maketitle
\abstract{The issue of whether we make decisions freely has vexed philosophers for millennia, Resolving this is vital for solving a diverse range of problems, from the physiology of how the brain makes decisions (and how we assign moral responsibility to those decisions) to the interpretation of experiments on entangled quantum particles. A deterministic model of free will is developed, based on two concepts. The first generalises the notion of initialisation of nonlinear systems where information cascades upscale from the Planck scale, exemplified by the chaology of colliding billiard balls, and featured in the author's Rational Quantum Mechanics. With `just-in-time' initialisation, such Planck-scale information is only initialised when it is needed to describe super-Planck scale evolution, and not e.g., at the time of the Big Bang. In this way determinism does not imply predestination and a system with finitely many degrees of freedom can shadow a system with infinitely many, over arbitrarily long timescales. The second concept describes the upscale control of such Planck-scale information on super-Planck scales and is illustrated by reference to stochastic rounding in numerical analysis. Using these concepts, a deterministic model is proposed whereby freely-made decisions are made by using past experiences to control the impact of noise in the low-energy brain. It is claimed that such a model has evolutionary advantages, not least preventing paralysis by analysis and encouraging rational risk taking. It is concluded that humans have free will, determinism notwithstanding. The model is applied to study the foundational issue of free choice in quantum physics experiments: it is shown that violating the Measurement Independence assumption does not invalidate the free-will conclusion above.}

\section{Introduction}

The question of whether we make decisions freely has vexed philosophers (indeed mankind) for millennia \cite{Kane}. The answer is not settled today. In a world governed by deterministic laws, where our current actions are determined by earlier states of the world, it seems everything we do is predestined by cosmological initial conditions \cite{Sapolsky}. This has troubling implications. Do we absolve Hitler from moral responsibility because his genocidal actions were predestined by initial conditions over which he manifestly had no control? Most of us surely recoil in disgust at such a thought. On the other hand, the question of moral responsibility is no more answered if our actions are the result of random neuronal fluctuations, over which we also have no control \cite{Harris}. 

This is not just the domain of philosophy. The notion of free will plays a key role in the interpretation of experiments on entangled qubits, such as Bell's experiment \cite{Brunner}. In particular, to prove a locally causal hidden-variable model of quantum physics satisfies the experimentally violated Bell inequality, one must assume that
\be
\label{MI}
\rho(\lambda | O_A, O_B)=\rho(\lambda)
\ee
where $\rho$ is a probability density, $\lambda$ is a so-called hidden variable describing properties of a particle (or pair of entangled particles) being measured, and $O_A \in \{0,1\}$ and $O_B \in \{0,1\}$ denote binary measurement settings under the control of independent experimenters `Alice' and 'Bob'. (\ref{MI}), known as the Measurement Independence assumption is typically interpreted as implying that experimenters are free to set their measuring apparatuses as they like, independent of the particles they measure. Indeed, (\ref{MI}) is often referred to as the `free choice' assumption e.g. \cite{Blasiak}. 

Free will is seen as vital for doing science. As Nobel laureate Anton Zeilinger put it:
\begin{quote}
The second important property of the world that we always implicitly assume is the freedom of the individual experimentalist.This is the assumption of 'free-will.' It is a free decision what measurement one wants to perform... This fundamental assumption is essential to doing science. If this were not true, then, I suggest it would make no sense at all to ask nature questions in an experiment, since then nature could determine what our questions are, and that could guide our questions such that we arrive at a false picture of nature.
\end{quote} 
This has left physicists with a quandary which has yet to be resolved: what is the key message behind the experimental violation of Bell's inequality?

The aim of this paper is threefold: firstly to propose a model of free will consistent with determinism, secondly to extend this to describe how we humans make decisions and assign moral responsibility to those decisions, and thirdly to propose that  the violation of (\ref{MI}) is not inconsistent with experimenter free will. In this latter respect, this paper is companion to the author's paper describing Rational Quantum Mechanics (RaQM) \cite{Palmer:2025a}. 

A starting point for the analysis in this paper is a quote from John Bell \cite{Bellb}
\begin{quote}
In this matter of causality it is a great inconvenience that the real world is given to us once only. We cannot know what would have happened if something had been different. We cannot repeat an experiment changing just one variable; the hands of the clock will have moved, and the moons of Jupiter. Physical theories are more amenable in this respect. We can calculate the consequences of changing free elements in a theory, be they only initial conditions, and so can explore the causal structure of the theory. 
\end{quote}
From this, we infer that: a) initial conditions are a specific instance of the more general concept of `free elements' of a physical theory, b) that describing the free elements of a physical theory is central to understanding the physical properties of that theory, and c) that the consistency of counterfactual worlds in a physical theory is important in determining that theory's causal structure. 

Many nonlinear dynamical systems in classical physics are governed by the butterfly effect as it was originally meant: whereby small-scale information can propagate upscale, influencing larger-scale information \cite{Lorenz:1969} \cite{Palmer:2014b}. Turbulence in classical fluids provides a familiar example of the butterfly effect. In Section \ref{bill} we describe a simple physical system based on the chaology of billiard balls \cite{Raymond} \cite{Berry:1988} to describe such an upscale cascade, making the point that the evolution of macroscopic scales may be generically and continuously sensitive to the upscale cascade of information from sub-Planck scales. In Sections \ref{bitshift} and \ref{padic}, we describe the conventional shift map and its $p$-adic generalisation, to describe upscale information propagation more quantitatively. in Section \ref{jit}, this shift map forms the basis for describing  the notion of `just-in-time' initialisation, where Planck-scale information is initialised only when it is needed for evolving super-Planck-scale information. With just-in-time initialisation, a finite-dimensional system can shadow an infinite-dimensional system over arbitrarily long periods of time, and, importantly, determinism does not imply predestination. Motivated in part by quantum physics, and in part by stochastic rounding in numerical analysis, in Section \ref{noise} we describe a modification of this shift map to describe an ability to partially control the impact of the inherently unpredictable information being injected to super-Planck scales arising from just-in-time initialisation. The concepts of just-in-time initialisation and impact control provide the basis for our model of human decision making and free will.  Section \ref{dfm} describes the application of the model to idealised and real-life situations. In Section \ref{quantum} we apply the model to quantum experiments (and hence to (\ref{MI})), where measurement settings have been freely chosen. 


\section{Upscale Cascades of Information in Nonlinear Deterministic Dynamics}
\label{nonlinear}

In this Section we develop the deterministic model for human decision making, free will and moral responsibility. If the reader becomes overwhelmed with technical details, it is recommended to skip to the next Section, which describes some application of the model, and in this way provides some motivation for these technical details.   

\subsection{Upscale Cascade of Information in Billiard-Ball Collisions}
\label{bill}

As mentioned, many nonlinear dynamical systems are governed by the butterfly effect, in the sense that the phrase was originally meant \cite{Lorenz:1969}, \cite{Palmer:2014b}: energetically insignificant small-scale information (e.g. flaps of butterflies' wings) propagating upscale, influencing energetically significant larger-scale information (e.g. the motion of weather systems). 

A simple physical example of such upscale cascade is that of colliding billiard balls \cite{Raymond} \cite{Berry:1988}. Let the radius of a billiard ball be $R$ and the mean free path between collisions $l$. The uncertainty $\Delta \theta_N$ in the direction of a billiard ball after $N$ collisions with other billiard balls, due to some very small initial angular uncertainty $\Delta \theta_0 \ll 1$, is given by  
\be
\Delta \theta_N \approx (\frac{l}{R})^N \Delta \theta_0
\ee
To get a feel for the power of such exponential growth, Berry \cite{Berry:1988} asks after how many collisions $N$ with other balls, the direction of motion of a billiard ball has been rendered completely uncertain ($\Delta \theta_N \approx 1$) by the gravitational field associated with a single electron at the edge of the observable universe? The answer, remarkably, is around 50. 

Here we describe a complementary sensitivity. The uncertainty $\Delta x_0$ in the position of the ball just before the first collision is equal to $l \Delta \theta_0$. After how many collisions $N$ will a positional uncertainty $\Delta x_0$ equal to the Planck length ($(\hbar G/c^{5})^{1/2} \approx 10^{-35}$m) induce complete uncertainty in the direction of motion of a ball, and hence its position on the billiard table. An order of magnitude estimation gives $N \approx 16$.  If instead of billiard ball collisions we consider colliding molecules in gas at standard pressure, then 16 collisions will occur quicker than the blink of an eye. 

In the absence of a theory of quantum gravity, we will assume that the theory which ultimately synthesises quantum and gravitational physics is itself a nonlinear theory:;not least the quantum measurement problem appears nonlinear. As such, motivated by the billiard ball example, we will assume sub-Planck-scale information is continually propagating upscale to influence the evolution of super-Planck classical scales, with quantum physics as an important intermediary as described in Section \ref{noise}. Such upscale propagation is a feature of RaQM \cite{Palmer:2025a}, a finite theory of quantum physics based on a discretisation of Hilbert Space. 

\subsection{Logistic and Bit-Shift Maps}
\label{bitshift}

In the classic initial value problem, we seek the solution $Z(t) \in \vmathbb R^K$, $t \ge t_0$ of a deterministic system
\be
\label{det}
\dot Z = f[Z] 
\ee
with initial conditions $Z(t_0)$. A key result in dynamical systems theory \cite{Strogatz} is that if $f$ is infinitely differentiable, the initial value problem has a unique solution $Z(t)$ over any time interval $[t_0, t]$. This is the mathematical basis for the notion of predestination and hence denial of free will: if the laws of physics are of the form (\ref{det}), then everything that occurs is uniquely predestined by (\ref{det}) and $Z(t_0)$, neither of which we have control over.  

Since the purpose of this paper is to be expository, we replace (\ref{det}) with a simple finite time-stepping scheme
\be
Z(t_{n+1})=F[Z(t_n)].
\ee
a chaotic example of which is the logistic map $F[Z]= 4(Z^2-Z)$ where $Z \in [0,1]$. In this paper, we develop ideas using the bit shift, topologically equivalent to the logistic map:  
\begin{align}
\label{bitshift1}
B: [0,1] &\rightarrow [0,1]  \nonumber \\
Z(t_{n+1})= B[Z(t_n)]&\equiv 2 Z(t_n) \pmod{1}
\end{align}
Writing out the base-2 expansion of $Z(t_n)$ explicitly,
\begin{align}
\label{bitshift2}
B: [0,1] &\rightarrow [0,1]  \nonumber \\
.X_n \;X_{n+1}\;X_{n+2}\;X_{n+3}\;X_{n+4}\; X_{n+5} \ldots. &\mapsto .X_{n+1}\;X_{n+2}\;X_{n+3}\;X_{n+4}\;X_{n+5}\;X_{n+6} \ldots
\end{align}
where $X_n \in \{0,1\}$. (\ref{bitshift1}) and (\ref{bitshift2}) have the continuity property discussed above. 

\subsection{The $p$-adic Shift Map}
\label{padic}

\begin{figure}
\centering
\includegraphics[scale=0.5]{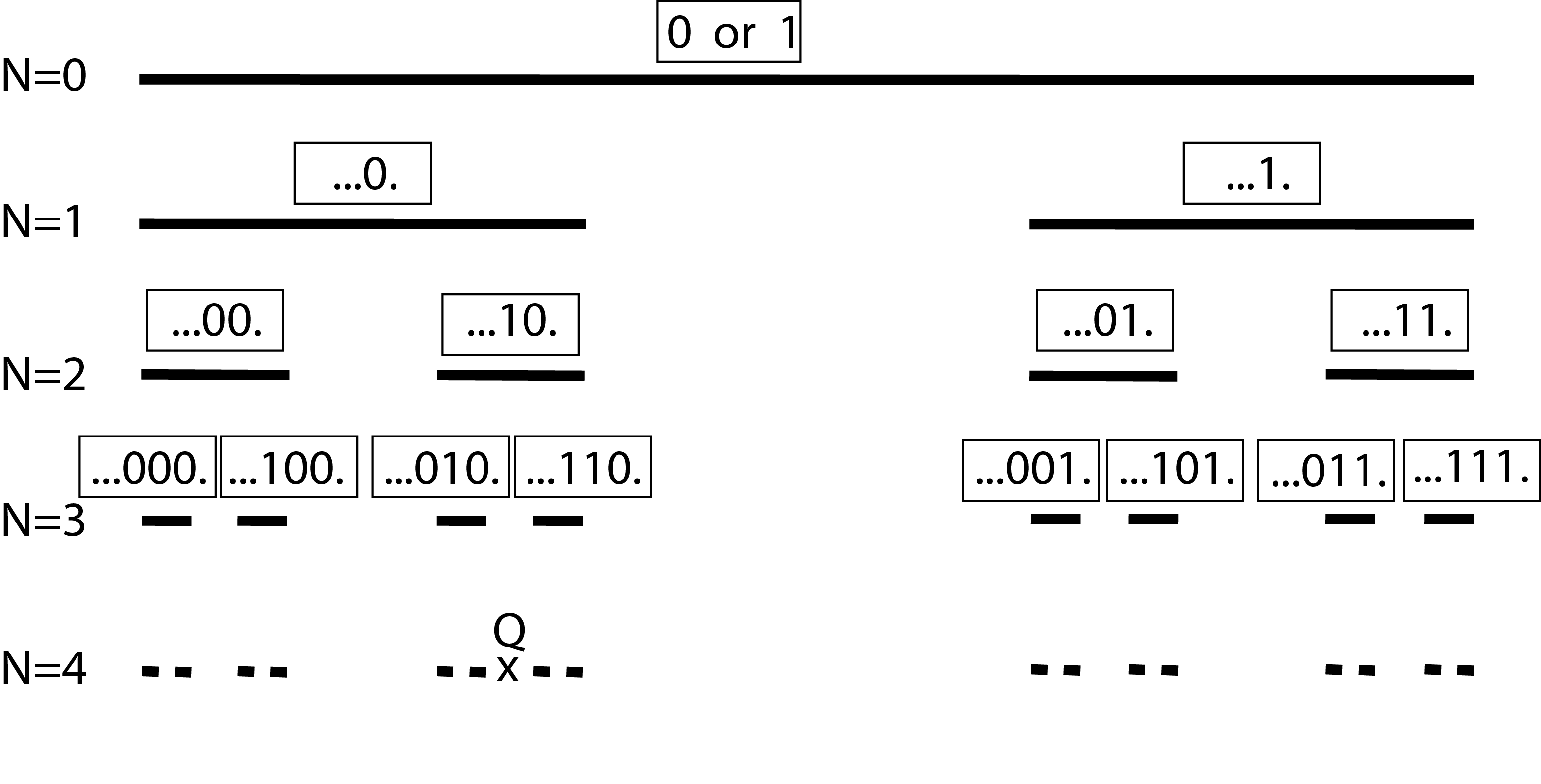}
\caption{\emph{Showing iterates of the Cantor set $C_2$, the pieces labelled by truncations of 2-adic integers. The figure also illustrates a hypothetical counterfactual state $Q$ not lying on $C_2$. The 2-adic distance between $Q$ and points on $C_2$ is undefined and hence, importantly, never small (even though the Euclidean distance between $Q$ and $C_2$ may appear small. }}
\label{cantor}
\end{figure}

The bit shift (\ref{bitshift2}) describes the evolution of a nonlinear system which is sensitive to small amplitude perturbations. However, in this paper we wish to describe systems which are sensitive to small-scale perturbations. We introduce the notion of scale by considering iterations of  the Cantor Set  $C_2=\cap_j C_2(j)$ (see Fig \ref{cantor}) - $C_2$ is in any case is the basis for RaQM \cite{Palmer:2025a}. The larger is $j$, the smaller in scale are the pieces of the $j$th iteration $C_2(j)$. It is well known (e.g. \cite{Katok}) that $C_2(j)$ can be labelled using 2-adic integers (see Fig \ref{cantor}) reflecting the homeomorphism between $C_2$ and the set $\vmathbb Z_2$ of 2-adic integers. The larger is $j$, the smaller is the 2-adic distance between the 2-adic integers that represent these neighbouring pieces. In this sense, although $p$-adic distance is unintuitive from a number-theoretic perspective, it is \emph{very} intuitive from this fractal geometric perspective. 

The $2$-adic counterpart of the bit shift (\ref{bitshift2}) can be written
\begin{align}
\label{2adic}
B_2: \vmathbb Z_2 &\rightarrow \vmathbb Z_2 \nonumber \\
\ldots  X_{5}\;X_{4}\;X_{3}\;X_{2}\; X_{1}\; X_0. &\mapsto \ldots X_{6}\;X_{5}\;X_{4}\;X_{3}\;X_{2}\;X_{1}.
\end{align}
where $\ldots  X_{5}X_{4}X_{3}X_{2}X_{1}X_0.$ denotes the base-2 representation of a 2-adic integer, and $X_i \in \{0,1\}$ as before. For those unfamiliar with $p$-adic numbers, one can simply think of the binary expansion of a 2-adic number as the mirror image of the binary expansion of a real. In particular, small-scale components of the state lie on the left and are shifted to the right in (\ref{2adic}), whilst for real numbers, small-amplitude components lie on the right and shift to the left in (\ref{bitshift2}). Using the 2-adic representation, it is possible to do basic arithmetic (and indeed analysis) on the Cantor set.  As such, (\ref{2adic}) is equivalent to a geometric zoom into the Cantor Set by a single fractal iteration. In RaQM, the passage of time $t_n \rightarrow t_{n+1}$ can be interpreted as a fractal zoom \cite{Palmer:2025a}. 

Using Fermat's Little Theorem with $p=2$, (\ref{2adic}) can be shown to be topologically equivalent to the generalised logistic map
\begin{align}
L_2: \vmathbb Z_2 &\rightarrow \vmathbb Z_2 \nonumber \\
Z &\mapsto \frac{1}{2} (Z^2-Z)
\end{align}
\cite{WoodcockSmart}.  

\subsection{All-at-Once vs Just-in-Time Initialisation}
\label{jit}

We can now introduce the first element of our deterministic model of free will. 

As discussed in Section \ref{bill}, we assume that sub-Planck information is generically and continually propagating upscale to influence the super-Planck classical scales (with conventional quantum physics as an important intermediary in a sense to be described in Section \ref{noise}). As such, we write the state of the $2$-adic shift map at some arbitrary time $t_n$ as
\be
\label{z2}
Z_f(t_n)=\underbrace{\ldots  X_{n+19}\;X_{n+18}\;X_{n+17}}_{\mathrm{sub-Planck}}\;\underbrace{X_{n+16}\; X_{n+15}\ldots X_{n+2}\; X_{n+1}\; X_n}_{\mathrm{super-Planck}}.
\ee
where $X_n$, $n>16$, represent scales smaller than the Planck length and $X_n$, $n \le 16$, represent scales larger than the Planck length. (The number 16, whilst motivated by the billiard ball example, is arbitrary - but see later in the section). 


The initial conditions for (\ref{z2}) can be written:
\be
\label{init}
Z(t_0)= \underbrace{\ldots X_{19} \; X_{18} \; X_{17}\; X_{16}\; X_{15} \ldots X_2 \; X_1\; X_0.}_{\mathrm{all\ at\ once}} 
\ee
where the phrase `all at once' denotes the conventional concept of initialisation: all components of $Z(t_0)$ (including, here, the infinitely many sub-Planck scales) are initialised simultaneously at $t_0$. 

Can a finite deterministic model, e.g., based only on the finitely many super-Planck scales, shadow (\ref{z2}) for arbitrarily long times $t_n$? At first sight it would appear this is impossible. If we truncate $X(t_0)$ to super-Planck scales, i.e., by 
\be
\label{finite}
Z_f(t_0)=\ X_{16} \; X_{15} \ldots X_2 \; X_1\; X_0.
\ee
then after $n>16$ timesteps
\be
Z_f(t_n)= \  0\;0\; \ldots 0.
\ee
i.e. the finite system has evolved to the fixed point $Z_f(t_n)=0$ and thereafter fails to shadow the chaotic $Z(t_n)$ (evolving by (\ref{2adic})) for all future times. 

However, let us now recall Bell's notion of initial conditions as merely a specific representation of the more general notion of `degrees of freedom' of a physical theory. With this in mind, consider again the finite model (\ref{finite}) of super-Planck scales, but now including the largest of the sub-Planck scales. We will assume that such sub-Planck information is initialised just before the timestep it is needed to determine the evolution of the super-Planck scales (and not earlier). For example, a sub-Planck component such as $X_{19}$ in (\ref{init}) plays no role \emph{at all} in determining the evolution of the super-Planck scales, until the shift map has been applied 3 times, whereon $X_{19}$ will take the 16th place in the state $X(t_3)$, and hence become a super-Planck component. 

\begin{figure}
\centering
\includegraphics[scale=0.5]{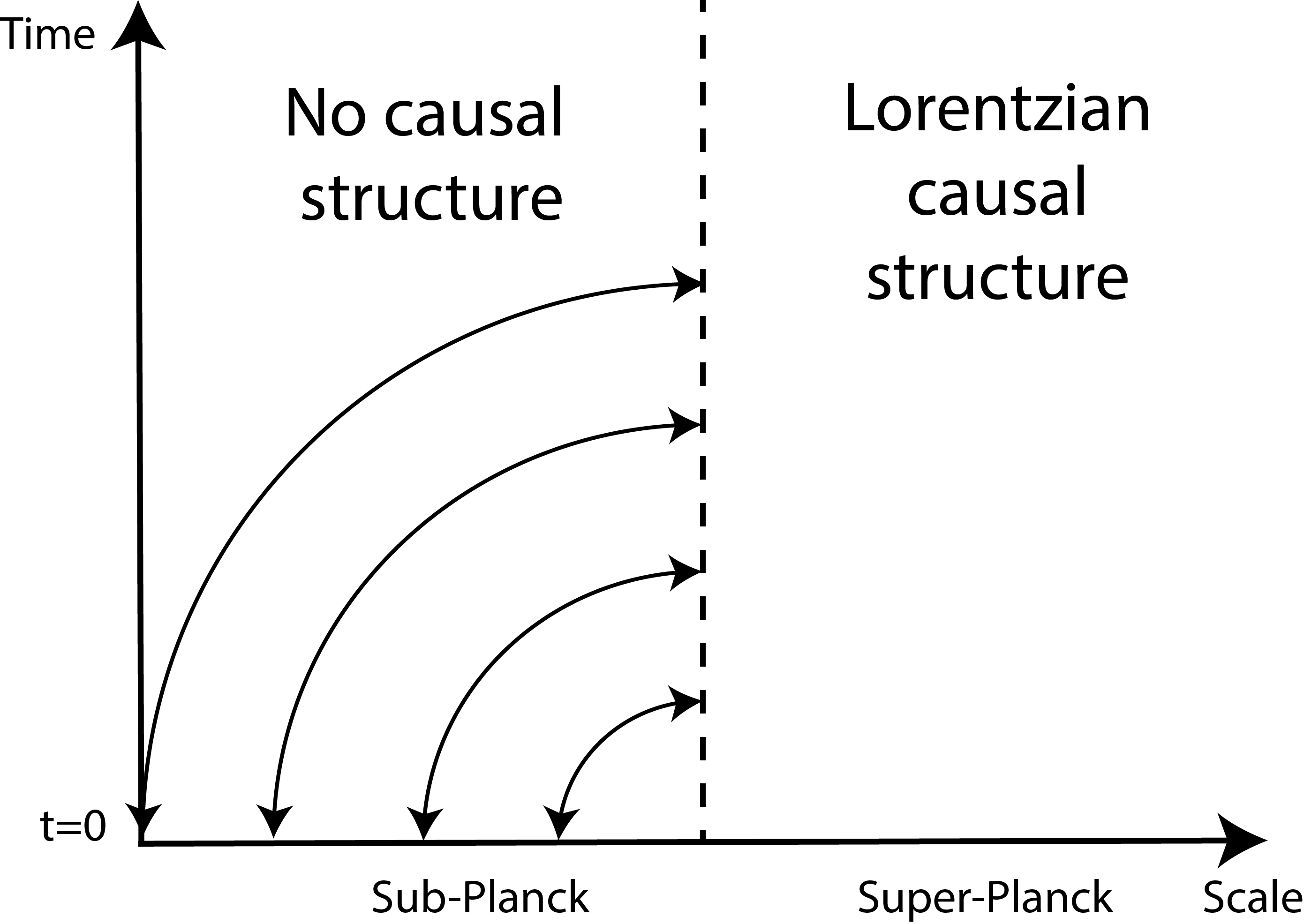}
\caption{\emph{Sub-Planckian scales are fundamentally inaccessible to super-Planckian scales and the Wheeler-DeWitt equation of quantum gravity makes no explicit reference to time. Reflecting this, we suppose that the causal structure of space-time is only manifest on super-Planck scales. As such, as far as the universe of super-Planck scales is concerned, there is no difference between a theory of the universe where sub-Planck scales are initialised at the time of the Big Bang, or are initialised just when they are needed, i.e., `just in time', to advance evolution of the super-Planck scales. From the latter perspective, sub-Planck information is fed to super-Planck scales as a boundary condition.}}
\label{planck}
\end{figure}

With 'just-in-time' initialisation of the sub-Planck scales (Fig \ref{planck}), temporal evolution in our finite shift map of super-Planck scales is given by
\begin{align}
\label{q}
Z_f(t_0)&=\underbrace{X_{17}}_{\mathrm{just\;in\;time}} X_{16} \ X_{15} \ldots X_2 \ X_1\ X_0.  \nonumber \\
Z_f(t_1)&=\underbrace{X_{18}}_{\mathrm{just\;in\;time}} X_{17} \  X_{16} \ldots X_3 \  X_2\ X_1.  \nonumber \\
\vdots \nonumber \\
Z_f(t_n)&=\underbrace{X_{n+17}}_{\mathrm{just\ in\ time}} X_{n+16} \ X_{n+15} \ldots X_{n+2}\  X_{n+1}\ X_n. \nonumber \\
\end{align}
Here, all the super-Planck scale components of the state are initialised conventionally, i.e. `all at once' at $t_0$ (consistent with the causal structure of space-time - see below). As far as the super-Planck scales are concerned, there is \emph{no difference} between the evolution of the infinite-component state (\ref{z2}) with all-at-once initialisation, and the evolution of the finite state (\ref{q}) with just-in-time initialisation. That is to say, with just-in-time initialisation, the finite system can shadow the infinite system arbitrarily far into the future (i.e. for arbitrarily large $n$). 

Just-in-time initialisation of sub-Planck scales makes physical sense on at least four grounds. The first is that sub-Planck length scales are inherently unknowable \cite{Mead} \cite{Hossenfelder}. For example, the wavelength of a photon needed to measure the position of an object on a scale smaller than the Planck length would have to be so large that the photon's self-gravitation would cause it to collapse into a black hole. It therefore makes no physical sense to assume the sub-Planck scales have the same ontological status as the super-Planck scales, as they do in (\ref{z2}). Secondly, just-in-time initialisation is consistent with the timeless and hence acausal nature of the Wheeler-DeWitt equation (of quantum gravity). What we refer to as space-time has a light-cone structure from which we can infer when one event is in the causal past, or causal future, of another. However, the Wheeler-DeWitt equation has no explicit reference to time. Based on this, one can surmise that causal structure does not exist for sub-Planckian scales, and its existence for space-time (on super-Planck scales) is an emergent property of the laws of physics. Consistent with this, the notion of an initial state in the causal past of some evolved state only applies to the super-Planck scales. Thirdly, as has already been noted, with just-in-time initialisation we can shadow the evolution of a system with infinitely many components, by a system with finitely many components. If one believes in the ultimate finiteness of the laws of physics, just-in-time initialisation makes good physical sense. Finally, it would seem profligate of the laws of physics to initialise variables before they are needed. See Fig \ref{planck}. 

Note that the finite shift map (\ref{q}) with just-in-time initialisation is as deterministic as the infinite shift map (\ref{z2}) with all-at-once initialisation. We have modified the notion of initialisation but not of determinism. Following Bell, we are merely defining the degrees of freedom of our dynamical system in a different way to the conventional approach. This is a critically important feature of our deterministic theory of free will: determinism is no longer synonymous with Big-Bang predestination. 

\subsection{Controlling the Impact of Sub-Planck Information}
\label{noise}

Just-in-time initialisation is the first important element in the proposed deterministic model for decision making, free will and moral responsibility. The second is an ability to control the upscale cascade of information from the sub-Planck scales. Before describing this at a technical level, it is worth giving a couple of examples of noise control. 

\begin{figure}
\centering
\includegraphics[scale=0.8]{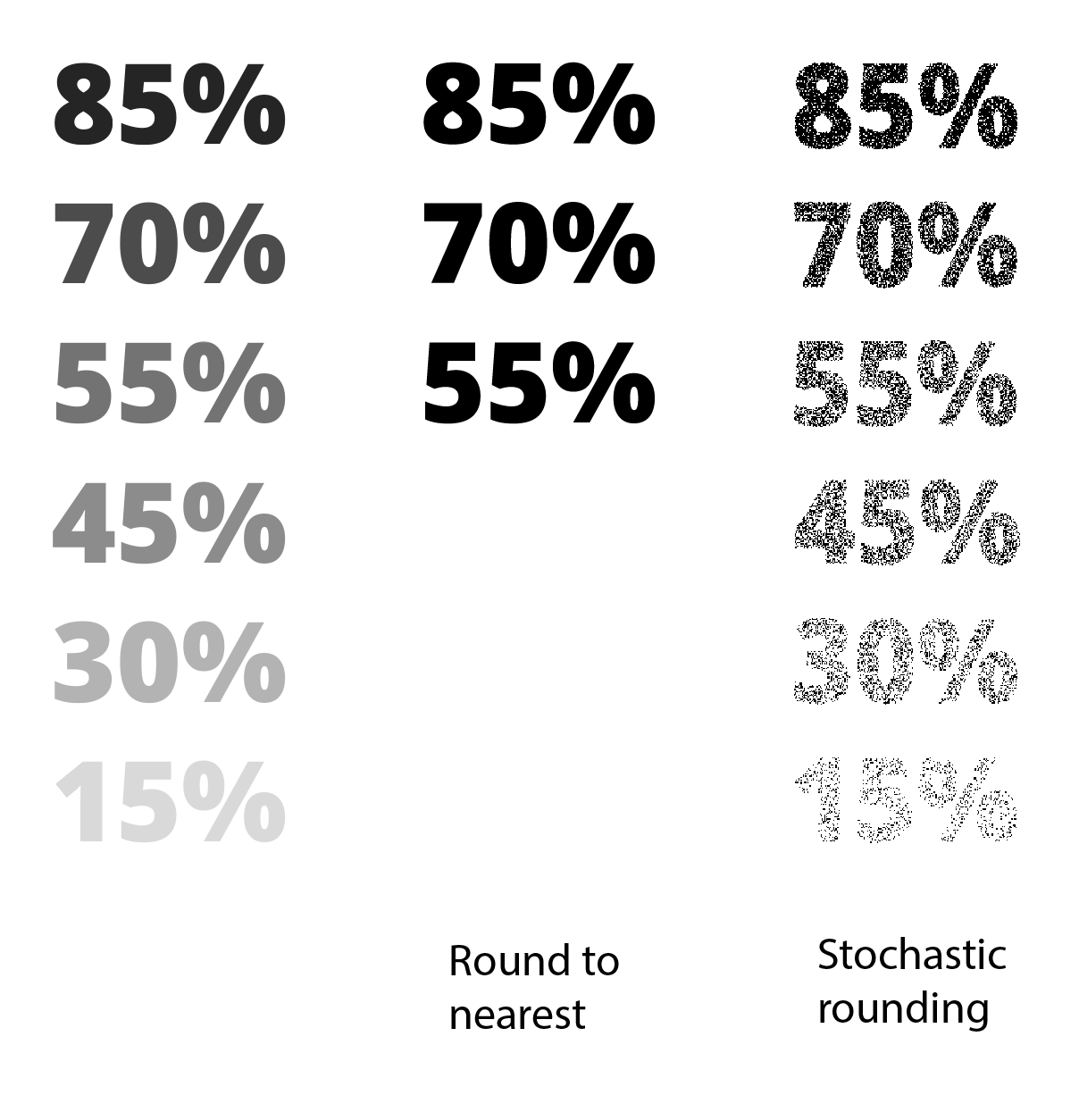}
\caption{\emph{In the left hand column the degree of greyness decreases smoothly. In the middle column the degree of greyness is rounded up to full black or down to full white according to a conventional `round-to-nearest' methodology. In the right-hand column, the degree of greyness is rounded up to full black or down to full white, according to a `stochastic rounding' methodology. Stochastic rounding may be utilised in the brain to minimise data transport between neurons. Here we argue it plays a central role in decision making.}}
\label{rounding}
\end{figure}

The first is a classical example: stochastic rounding in numerical analysis \cite{Hopkins}. This is an important technique to allow numerical models to solve equations of the form (\ref{det}) at low numerical precision, and hence to allow computations to be performed with relatively small expenditure of physical energy. Here random numbers are used to arrive at the numerical solution, but the impact of such numbers is controlled by the size of the truncation error. Suppose $Z_a < Z < Z_b$ where $Z_a$ and $Z_b$ are the two numbers which lie closest to $Z$ in the set of numbers representable in reduced-precision format. Instead of systematically rounding to the nearest (here nearer) representable number, as in traditional numerical analysis, $Z$ is rounded to $Z_a$ with probability $\alpha=(Z_b-Z)/(Z_b-Z_a)$ and to $Z_b$ with probability $1-\alpha=(Z-Z_a)/(Z_b-Z_a)$. Hence, when $Z$ is close to $Z_a$, the noise has almost no impact and it is very likely that $Z$ will be rounded to $Z_a$. Similarly, when $Z$ is close to $Z_b$, it is very likely that $Z$ will be rounded to $Z_b$. However, when $Z$ is equidistant between $Z_a$ and $Z_b$, then the noise has maximal impact and it is as likely that $Z$ will be rounded to $Z_a$ as to $Z_b$. Here $\alpha$ is a parameter which controls the impact of the noise in the numerical solution. An example of the value of stochastic rounding is illustrated in Fig \ref{rounding}. Stochastic rounding helps alleviate the problem of accumulating round-off error compared with the more traditional `round to nearest' strategy. 

\begin{figure}
\centering
\includegraphics[scale=0.5]{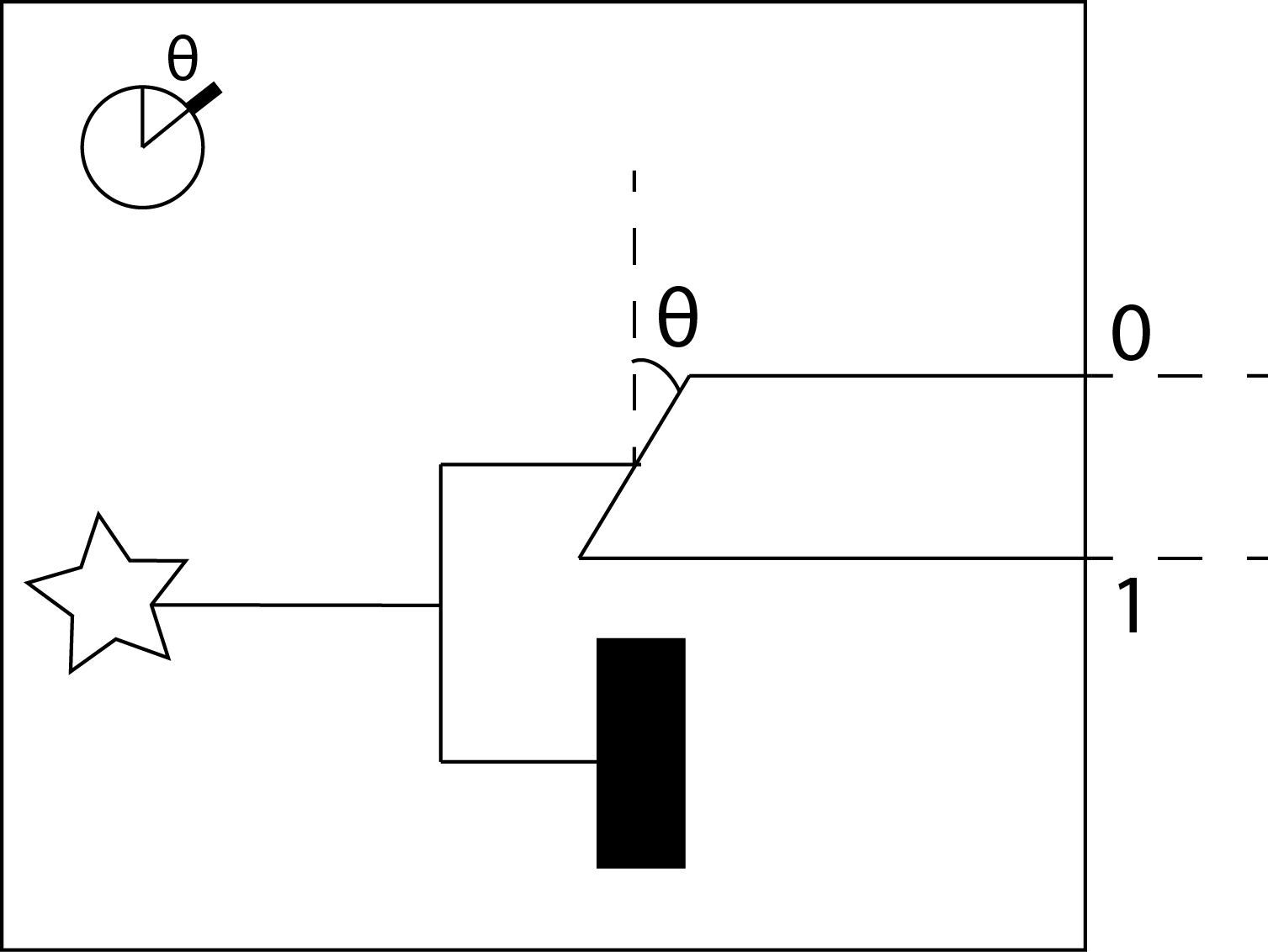}
\caption{\emph{Schematic figure of how the experimenter can control the degree of apparent randomness from a quantum source of particles. Here 0 refers to a `spin-up' channel and 1 to `spin down'}}
\label{SG}
\end{figure}

The second utilises quantum physics. Fig \ref{SG} shows (very schematically) a source of spin-1/2 particles being fed into a Stern-Gerlach device oriented in a known reference direction. The spin-up output beam is fed into a second Stern-Gerlach apparatus oriented at angle $\theta$ to the reference direction. An experimenter can adjust $\theta$ freely by turning a knob. When $\theta=90^\circ$ particles are outputted through the two output channels (0 and 1)  in random order. The closer $\theta$ is to $0^\circ$ ($180^\circ$), the more the particles are output through the 0 channel (1 channel respectively). Using the knob, the experimenter can control the probability of a particle being emitted through the 0 and 1 channels. 

To incorporate this notion of control into the shift map, the shift map is modified in two steps. First we write the state of the finite shift map as 
\be
\label{Z}
Z_f(t_n)=\underbrace{Y_{n+3}}_{\mathrm{jit}}\ \underbrace{Y_{n+2}}_{\mathrm{quant}} \ \underbrace{X_{n+1}\  X_n}_{\mathrm{classical}}.
\ee
Here the elements $X_i$ over the `classical' underbrace each correspond to bits 0 or 1 as before. For reasons to be discussed later in this section, we restrict these classical bits to just two. On the other hand, the elements $Y_i$ over the `jit' and `quant' underbraces are whole numbers whose binary expansion comprises $2^N$ bits e.g,
\be
Y_{n+2}=\underbrace{a\;b\;c \ldots d.}_{2^N\ \mathrm{bits}}
\ee
where $a$, $b$, $c$ $\in \{0,1\}$, and therefore takes values in $\{0,1,2 \ldots, 2^{2^N}-1\}$. (Here $N$ relates directly to its use as an ensemble size in RaQM \cite{Palmer:2025a}). Temporal evolution proceeds as the generalised shift map:  
\begin{align}
\label{mod}
Z(t_n)&=\underbrace{Y_{n+3}}_{\mathrm{jit}}\ \underbrace{Y_{n+2}}_{\mathrm{quant}} \ \underbrace{X_{n+1}\ X_n.}_{\mathrm{classical}} \nonumber \\
\mapsto Z(t_{n+1}) &= \underbrace{Y_{n+4}}_{\mathrm{jit}}\ \underbrace{Y_{n+3}}_{\mathrm{quant}} \ \underbrace{X_{n+2}\  X_{n+1}.}_{\mathrm{classical}} 
\end{align}
where 
\be
X_{n+2}=\mathscr D_{\alpha_n}(Y_{n+2})
\ee
$\mathscr D_{\alpha_n}$ is a deterministic `collapse' function which maps an ensemble of $2^N$ bits (typically $10^{200}$ for a qubit in a quantum computer according to RaQM), to $2^0=1$ classical bit.  $\alpha_n$ is a `control parameter' defined below.  

Consider first the case $N=1$. Then the binary expansion of $Y_{n}$ comprises $2^1$ bits which can be written $ab.\ $. The 4 possible values of $a$ and $b$ are given by the four columns of the table
\begin{center}
\be
\begin{array}{ccccc}
\label{bitst}
a \ =& 0 & 0 & 1 & 1  \\
b \ =& 0 & 1 & 0 & 1
\end{array}
\ee
\end{center}
For each of these 4 values, the 4 possible values of each of the following combination of bits are
\begin{center}
\be
\begin{array}{ccccc}
\label{bitst}
\ \ \ \ \ a \times a'=& 0 & 0 & 0 & 0  \\
\ \ \ \ \ a \times b =& 0 & 0 & 0 & 1  \\
\ \ \ \ \ \ \ \ \ \ a =& 0 & 0 & 1 & 1  \\
(a' \times b')'=& 0 & 1 & 1 & 1  \\
(a \times a')'=& 1 & 1 & 1 & 1 \\
\end{array}
\ee
\end{center}
where $a'$ is the bit complement of $a$ etc. 
Hence, if we assume that the values $a \in \{0,1\}$ and $b \in \{0,1\}$ are equally likely, then
\begin{align}
\label{prob}
P(a \times a'=1)&=0 \nonumber \\
P(a \times b=1)&=1/4 \nonumber \\
P(a =1)&=1/2 \nonumber \\
P((a' \times b')'=1)&=3/4 \nonumber \\
P( {(a \times a')'}=1)&=1
\end{align}
The deterministic function $\mathscr D_{\alpha_n}$ is therefore defined as:
\begin{align}
\label{Ds}
\mathscr D_{0}(Y_n)&=a \times a' \nonumber \\
\mathscr D_{1/4}(Y_n)&=a \times b \nonumber \\
\mathscr D_{1/2}(Y_n)&=a  \nonumber \\
\mathscr D_{3/4}(Y_n)&={(a' \times b')'} \nonumber \\
\mathscr D_{1}(Y_n)&= {(a \times a')'} 
\end{align}
From (\ref{prob}), with $X_n = \mathscr D_{\alpha_n}(Y_n)$ and $\alpha_n \in \{0, 1/4, 1/2, 3/4, 1\}$
\be
\label{prob2}
P \left(\mathscr D_{\alpha_n} (Y_n)=1 \right) = \alpha_n
\ee
In this way, specifying a value $\alpha_n=0$ or $1$ implies that  the output $X_n$ does not depend on the input $Y_n$. If we have control over $\alpha_n$, then applying a value $\alpha_n=0$ or $1$ implies complete control on the output $X_n$. By contrast, applying a value $\alpha_n=1/2$ indicates we have have no control on the output $X_n$, which is instead entirely dependent on the bit $a$ and hence the input $Y_n$. For other values of $\alpha_n$ we have partial but not complete control over the output $X_n$. 

This construction is easily generalised. For $N=2$, $Y_n$ can be written in the bit form $abcd.$ i.e., using $2^2$ bits. This quadruple of bits takes $2^{2^2}=16$ possible values. Using (\ref{bitst}) in block form, we can construct products of the four bits (e.g., $a\times b\times c \times d$) and their complements such that all probabilities between 0 and 1, in steps of 1/16th, are represented. Now (\ref{prob2}) holds with $\alpha_n \in \{0, 1/16, 1/8, \ldots, 15/16,1\}$. This procedure can be used to generate values of $\alpha_n$ between 0 and 1 in arbitrarily small steps. 

We now write  (\ref{Z}) as 
\be
\label{Z2}
Z_f(t_n)=\underbrace{Y_{n+3}}_{\mathrm{jit}}\ \underbrace{Y_{n+2}}_{\mathrm{quant}} \ \underbrace{X_{n+1}}_{\mathrm{decide}} \ \underbrace{X_{n}.}_{\mathrm{act}}
\ee
where the classical bit $X_{n+1}$ corresponds to deciding (or choosing), and the other classical bit $X_n$ corresponds to acting on that decision (or choice). We can interpret this short-chain of events as follows. At timestep $n$, ensemble information is just-in-time initialised into the largest sub-Planck scale. At timestep $n+1$, this information is transferred to the quantum world. At timestep $n+2$, under the action of the control function $\mathscr D_{\alpha_{n+2}}$, this ensemble information is transformed into a classical decision bit 0 or 1: decide to do something or not to do something. At timestep $n+3$, that decision becomes an action: do something or don't do something. The question of whether it is possible to encompass this upscale cascade with a single fixed timestep is addressed briefly in the Appendix. 

\section{Decision Making, Free Will and Moral Responsibility}
\label{dfm}

The model described above is now put to use, to formalise decision making, free will and moral responsibility from the deterministic perspective of just-in-time initialisation with control. 

\subsection{Decision Making}

How do we humans make decisions? It might seem that the best strategy is to weigh up the pros and cons of each option and decide on the one where the `pros minus cons' is largest. The problem is that pros and cons are typically not comparable. Suppose we have been offered a job overseas. How do we compare quantitatively the gain in salary against the fact that we will become distant from close family members? Perhaps on 40\% of occasions where we try to make a decision, we tentatively conclude that we should accept the offer (option 1), whilst on the other 60\% of occasions we conclude that we should reject the offer and stay put (option 0). What to do? 

The fact of the matter is that for all but the more trivial decisions of life, there is no maximum `pros minus cons' to compute. Indeed, trying to compute the elusive maximum leads to the classic `paralysis by analysis' which prevents a decision being made at all. In trying to do the impossible, we prevaricate and procrastinate. From an evolutionary perspective this is not a good strategy for survival - if we are about to be attacked by a predator, and have to decide between one of two escape routes, a moment's delay in weighing up the pros and cons could prove fatal: in such a situation, some decision is better than no decision.  

The human brain is a noisy organ (\cite{RollsDeco}), by virtue of the fact that the body only supplies 20W to power some 80 billion neurons. However, there is good evidence that the brain makes constructive use of such noise, e.g. making us humans the intelligent creative species we are \cite{PalmerOShea}. Here it is argued that noise in the brain plays a vital role in human decision making (arguably in other animals too). But we do not use the word `noise' here as something inherently random. Rather, its origin is the deterministic but completely unpredictable Planck-scale information which is being continually injected into the classical world, and hence into the functioning of the brain. 

 
Following the analysis in Section \ref{noise}, we propose that a decision process is associated with the injection from $Y_n \in \{0, 1, \ldots 2^{2^N}-1\}$ to $X_n=\mathscr D_{\alpha_n}(Y_n) \in \{0,1\}$, where $\alpha_n$ controls the probability that $X_n=1$, given no prior knowledge about $Y_n$ other than each of its binary digits is equally likely. As we discuss below, setting the value of $\alpha_n$ is a semi-conscious process, determined by the decision maker's previous decisions and their outcomes.

For example, consider the job-offer problem. As discussed, when analysing the job offer multiple times with the brain's slow methodical analytic skills (what Kahneman calls `System 2' \cite{Kahneman}), we conclude we would accept (Option 1) on 40\% of occasions, but reject on 60\% of occasions. But now the phone rings and it's time to make that decision for real. The proposal here is that past efforts to analyse the problem (more generally past experiences) lead to the control parameter $\alpha_n$ being set to a value around 0.4. That is, drawing on past experiences only permits weak control on the value $X_n=\mathscr D_{\alpha_n}(Y_n)$ (recall that $\alpha_n=0.5$ corresponds to no control). On the other hand, if those past efforts at analysis have suggested that we should almost certainly accept the offer, then perhaps $\alpha_n =0.9$ and nine times out of ten we would accept the offer. 

By analogy with stochastic rounding, we can call this the 'stochastic-rounding-with-control' strategy (though recognising that the input $Y_n$ only appears for practical purposes to be stochastic because it is inherently unpredictable). Hence, using stochastic rounding with control, in the first example, there is a 40\% probability that $X_n=\mathscr D_{0.4}(Y_n)=1$ (and hence that we will accept job offer) and hence a 60\% probability $X_n=\mathscr D_{0.4}(Y_i)\}=0$ (and hence that we will reject job offer). By contrast, in the second example, there is a 90\% probability that we will accept the job offer. 

Is this better than a `round-to-nearest' strategy, where the brain systematically rounds $\alpha_n =0.4$ to 0 and systematically round $\alpha_n=0.9$ to 1. We might postulate such a strategy on the basis that all decisions should be conscious decisions where we are fully in control. With the round-to-nearest strategy, then we will certainly reject the job offer in the first example, and certainly accept the job offer in the second example. 

There are a number of advantages to the stochastic-rounding-with control strategy over the round-to-nearest strategy. The first is related to the issue of accumulation of round-off error. in numerical analysis If we always `round to nearest' then we will be systematically opting for a risk averse strategy. Arguably one of the key reasons why humans have thrived as a species is by not being irrationally (i.e., pathologically) risk averse, i.e., by exploring opportunities when they arise, even when they are uncertain. That is to say, `round to nearest' is not a strategy consistent with our being the dominant species on the planet. This does not mean taking irrational risks, but, in an ensemble of situations where we estimate a 40\% probability of option 1 being better than option 0, it is rational to decide on Option 1 on 40\% of occasions (as opposed to never deciding on Option 1 with the round-to-nearest strategy).  

It is often said that successful entrepreneurs can make decisions very quickly, relying on their `gut instinct' - what Kahnaman would call `System 1' \cite{Kahneman}. By contrast, scientists tend to over-analyse situations and are poor at taking advantage of entrepreneurial opportunities. Perhaps the difference is that successful entrepreneurs take advantage of the brain's capability of making decisions by stochastic rounding with control, whilst scientists' over-developed `System 2' analytic skills override this capability and tend irrationally to round to nearest. Perhaps this is why scientists do not typically make good entrepreneurs. 

A second advantage of stochastic rounding with control is that instantaneous decisions can always be made when necessary (using Kahneman's fast System 1). By default, in lieu of any past experience, a decision process can always proceed with the default $\alpha_n=1/2$. Death through `paralysis by analysis' will never occur with the stochastic-rounding strategy. A third advantage is that stochastic rounding is a low-energy strategy. A decision can be made without having to devote large amounts of energy to estimating $\alpha_n$ to great precision (which may be impossible anyway). Indeed, for unimportant decisions, a rational decision strategy involves one where only small amounts of energy are expended estimating $\alpha_n$, and instead the principal source of information being used to make the decision is the noise itself. This may mean making a `wrong' decision on occasions when expenditure of large amounts of energy could lead to more `right' decisions. However, if the consequences of being right or wrong are utterly unimportant, it could be irrational to spend large amounts of energy being right more often than being wrong. Being wrong does not mean being irrational. 

In stochastic rounding with control, we refer back to past analyses, and to past decisions and their consequences, in estimating $\alpha_n$. As small children with few past experiences to fall back on, we default to $\alpha_n=1/2$ (though genetic preconditioning can presumably provide values $\alpha_n \ne 0.5$). However, a bad experience can lead to a rapid update in $\alpha_n$, e.g. when next deciding whether to dip one's hand into a bowl of steaming water. In this way, we learn the danger signs of steam. At the other end of life, many of us get `set in our ways' as we get older, so that $\alpha_n$ values naturally tend to either 0 or 1 as $n \rightarrow \infty$ and we lose the benefits of stochastic rounding with control, and default to the less risk averse round-to-nearest strategy. 

\subsection{Free Will}

In this section it is argued that the deterministic shift map (\ref{2adic}), with just-in-time initialisation and stochastic rounding with control, is compatible with free will: the capacity of humans to make decisions or perform actions which are not wholly dependent on prior events or states of the universe and are consistent with the decision-maker's desires. 

The case that free will is inconsistent with determinism has been put recently by Sapolsky \cite{Sapolsky}:
\begin{quote}
To reiterate, when you behave in a particular way, which is to say when your brain has generated a particular behavior, it is because of the determinism that came just before, which was caused by the determinism just before that, all the way down.....And when people claim that there are causeless causes of your behavior that they call `free will', they have (a) failed to recognize or not learned about the determinism lurking below the surface and/or (b) erroneously concluded that the rarefied aspects of the universe that do work indeterministically can explain your character, morals and behavior. 
\end{quote}
In the proposed model based on the shift map (\ref{2adic}), `all the way down' does not mean going back to the Big Bang. Rather it means going down to Planck scales. With just-in-time initialisation, such Planck scale information is not initialised at the time of the Big Bang. 

However, since these specific initialised variables are by their nature unpredictable, then for all practical purposes they are no different to noise. But then it would seem our actions are not preconditioned by our desires, i.e., by our will. To quote Harris \cite{Harris}. 
\begin{quote}
If my decision to have a second cup of coffee this morning was due to a random release of neurotransmitters, how could the indeterminacy of the initiating event count as the free exercise of my will? Chance occurrences are by definitions ones for which I can claim no responsibility. And if certain of my behaviors are truly the result of chance, they should be surprising \emph{even to me}. How would neurological ambushes of this kind make me free?
\end{quote}
However, Harris's example is simply a special case of the stochastic-rounding-with-control model discussed above, but where $\alpha_n=1/2$, i.e., without control. If we have chosen to apply no control to the input $Y_n$ (we couldn't care less one way or the other whether we have a second cup of coffee) we should not be surprised if the outcome $X_n$ is 0 or 1. However, if we really wanted a second cup of coffee (because in the past we have tended to feel better after two cups of coffee), and consequently apply, say, an $\alpha_n =0.9$ to the decision process, then we should be surprised if we were to decide not to have that second cup of coffee. Perhaps on the odd (one in ten) occasion when we do decide not to have the second cup, we'll find a way of justifying that decision (e.g., it'll only make me want to pee, and I am due to go out in half an hour). After all, there must have been some reason why $\alpha_n \ne 1$. 

One famous experiment that is commonly brought up in any modern discussion of free will is that by the physiologist Benjamin Libert \cite{Libet} using EEG to show that activity in the brain's cortex can be detected some 300 milliseconds before a person feels that they have decided to move. This notion indicating that a decision has been made earlier than the time the subject is aware of that decision, has now been confirmed many times. As Harris puts it:
\begin{quote}
One fact now seems indisputable: Some moments before you are aware of what you will do next - a time in which you subjectively appear to have complete freedom to behave however you please - your brain has already determined what you will do. You then become conscious of this `decision' and believe that you are in the process of making it.
\end{quote}

This phenomenon is entirely consistent with the model proposed in Section \ref{nonlinear}. As discussed, we input our past experiences into the decision process through the control parameter $\alpha_n$. In lieu of relevant past experiences, the decision process defaults to $\alpha_n = 1/2$ guaranteeing that we can make a decision quickly if needs be. Our estimate of $\alpha_n$ is not itself part of the decision process, it is a precursor to the decision process. In the example where we have been made a job offer, the brain's estimate $\alpha_n= 0.4$ is merely the result of a partially unconscious synthesis of earlier analysis. The evaluation of the function $\mathscr D_{\alpha_n}(Y_n)$ is itself a neurological process applying the estimated value of $\alpha_n$ to an instance $Y_n$ of the neuronal noisy input The result $X_n$ is communicated to our conscious thought some hundreds of milliseconds later. That is to say, the conscious (or semi-conscious) process in decision making is in estimating $\alpha_n$. It is not in executing the injection $X_n=\mathscr D_{\alpha_{n}}(Y_n)$ This is what Libert's experiment shows. 

For many, free will is defined by an ability to have done otherwise. By asserting `I could have done otherwise', we imagine a counterfactual world, the same as the real world in all respects except that if I decided $O=1$ in the real world, I decided $O=0$ in the counterfactual world. If initial conditions are set at the time of the Big Bang, it is hard to imagine what initial perturbation to the Big Bang could have this localised impact. A typical small perturbation at initial time will have dispersed over all degrees of freedom $10^{10}$ years later, affecting only my decision but everything else in the causal future of that perturbation. As such, a small initial perturbation leading to a highly localised perturbation some $10^{10}$ years later will be incredibly fine tuned, if it can be shown to exist at all. However, with just-in-time initialisation, the shift model (\ref{z2}) illustrates precisely how to target decision processes to specific initialised variables. By changing a specific just-in-time initialisation $Y_{n}$ (itself a free variable of the model by definition), we can counterfactually change the decision $\mathscr D_{\alpha_n} (Y_{n})$ keeping other components of the universe fixed (including the estimate $\alpha_n$. Such targeted counterfactual worlds are wholly consistent with just-in-time initialisation and stochastic rounding with control (though see Section \ref{quantum} for specific types of counterfactual worlds that are not consistent with the proposed deterministic model.) We do have free will.  

\subsection{Moral Responsibility}

Predestination makes the notion of moral responsibility meaningless. How can one be morally responsible for actions that depend on data initialised at the dawn of time? By contrast, the deterministic model presented here allows us to define moral responsibility consistent with common-sense ideas. We illustrate the stochastic-rounding-with-control strategy with examples presented by Harris \cite{Harris}. Here we are considering circumstances in which a decision is made to fire a gun pointed at someone (Option 1). Society considers this a punishable crime under normal circumstances. We can frame the question of moral responsibility in terms of whether we believe the decision was made with an $\alpha_n$ substantially less than unity (let's say $<0.9$). 
\begin{itemize}
\item
A four-year-old boy was playing with his father's gun and killed a young woman. The gun had been kept loaded and unsecured in a dresser drawer.
\item 
A 12-year-old boy who had been the victim of continual physical and emotional abuse took his father's gun and intentionally killed a young woman because she was teasing him. 
\item
A 25-year-old man who had been the victim of continual abuse as a child intentionally shot and killed his girlfriend because she left him for another man. 
\item
A 25-year-old man who had been raised by wonderful parents and never abused, intentionally shot and killed a young woman he had never met `just for the fun of it'.
\item
As in the item above, but an MRI of the man's brain revealed a tumor the size of a golf ball in his medial prefrontal cortex.  
\end{itemize}
In the first example, the child had not accumulated enough experiences to know that guns were deadly. His decision to pull the trigger was based on a value $\alpha_n \approx 1$ formed by a curiosity to get to know the world around him better - in the past such curiosity has served the boy well. In the second example, the young boy similarly has not accumulated sufficient past experiences to form a sound judgement on the circumstances when use of a gun can be justified. As in the first example, a value $\alpha_n \approx 1$ should be viewed merely as indicative of immaturity. As a society, we judge that by the time a human has reached adulthood, he should, normally at least, have had accumulated sufficient experiences to make sound estimates of $\alpha_n$. The third example, indicates circumstances where prior circumstances lead to faulty estimates of $\alpha_n$, leading to a larger value than for an individual with a normal upbringing. Society may take this into account as mitigating circumstances when convicting the man, though nevertheless considers that the man should have known that killing his girlfriend was a crime and hence his $\alpha_n$ must, mitigating circumstances notwithstanding, have been substantially less than 1. We convict the man in the fourth example, not just to keep a dangerous individual off the street, but because there are no mitigating circumstances to suspect that his $\alpha_n$ was pathologically close to 1. That is to say, we judge that the man must have been aware of the negative consequences, to him if not to the family of the woman he shot. It is our belief that the man took the negative consequences into account in forming his value of $\alpha_n$ and that consequently this $\alpha_n$ was substantially less than 1, provides our judgement that he had moral responsibility for his actions and therefore should be punished. Of course, there may be pathological medical reasons why $\alpha_n \approx 1$ in an adult, as in the fifth example, in which case the man cannot be held morally responsible. 

In a conventional deterministic universe, it is argued that punishment should not be given in retribution for a crime, but firstly as a deterrent, not only to stop the criminal themselves from reoffending, but also for those who may be tempted themselves to commit a similar crime, and secondly to keep a dangerous individual off the streets where they may reoffend. Whilst these are reasons for punishment, they surely can't be the only reasons. If Hitler had been caught alive, there seems little doubt that he would have been executed. This would indeed have stopped him from committing genocide again. However, he was a busted flush and there was very little prospect of him getting back to power. And genocide is such a singularly perverse crime against humanity that executing Hitler is hardly likely to deter some future genocidal maniac. Indeed, perhaps such individuals would welcome the prospect of being executed, especially if their crimes were religiously motivated. On this basis, one might argue that it would have been enough, perhaps even better, to put Hitler under house arrest (e.g., as Napoleon was, on St Helena). But the vast majority of the population would have been utterly outraged by such a decision. If it was put to them that this is the logical consequence of living in a deterministic world, then they would respond, with conviction, that we therefore don't live in a deterministic world. Here we have proposed a different way of thinking of determinism, which can accommodate the intuition of most of us: it does make sense to hold criminals such as Hitler morally responsible and punish them, purely for the purposes of retribution. 

However, we finish with another recent real-world example, which illustrates that although the stochastic rounding strategy can be considered rational over the round to nearest strategy, there is no guarantee that it will lead to the right decision in any one instance. 

In 1994, following the unexpected death of his brother Bassel in a car accident, Bashar al Assad returned to Syria from the UK where he had been training as an ophthalmologist, helping his patients lead better lives \cite{BBC:2021}. He is now heir apparent to the presidency, and becomes president in 2000 following the death of his father. Influenced by values assimilated during his time in the UK, Assad presents himself as a liberal reformer and achieves a measure of popularity at home and abroad.  However, in 2001 Assad must decide whether to forcefully put down demonstrations in the city of Homs. The demonstrators are complaining about the torture of small boys by the mayor of Homs, a relation of Assad. In making this decision, Assad is influenced by his mother who advises that to give in to the demonstrators would be a sign of weakness and would at a stroke undo the hard work of his father. As an ophthalmologist, Assad has not been prepared for this type of decision. He decides to follow his  mother's advice and crack down on the demonstrators, with deadly consequences. This leads to general uprisings. Assad, is now caught in a deadly downward spiral, in which over half a million Syrian people are eventually killed, some by chemical weapons. Like Hitler before him, Assad is now viewed inside his country and internationally as a brutal genocidal murderer. 

Of course, we do not know what went on in Assad's head in the early days of his presidency. Nevertheless, one can imagine he must have been conflicted between the liberal values he experienced when living in London, with his mother's values based on the preeminent importance of upholding family honour. There is no evidence of some cognitive pathology that would have led to an $\alpha_n \approx 1$ (crack down on the demonstrators) in these early days. If he made the decision with some misgivings, perhaps a value in the range $0.4 \le \alpha_n \le 0.6$ might be accurate. However, the more Assad got caught in the  downward spiral of violence, the more $\alpha_n \rightarrow 1$ and the more pathological Assad's decisions became. 

As per the Harris example, this would be consistent with the notion that Assad had moral responsibility for his decisions, at least early on. Nevertheless, the example raises an uncomfortable issue. The values $\alpha_n$ are determined by past experiences. If any one of us, raised in a liberal democracy that values life above almost everything, were suddenly uprooted in this way, simply because of an unexpected death in the family, would these past experiences have prepared us sufficiently to ignore the insistence from family members, one's own mother in particular, to uphold family honour? Like the 4-year old in Harris's example, experiences in life have not prepared us for this situation.  This is an example of the human predicament, now \emph{in extremis}. Our experiences don't always prepare us properly for making the key decisions of our life. The brain has evolved a process precisely in order to deal with these situations: we utilise noise in the brain to make decisions when otherwise we might become paralysed by analysis, being unable to meaningfully compare apples (liberal values) with oranges (mother's insistence). But of course there is no guarantee that this strategy will not lead us to make specific decisions we ultimately regret, and which, by any decent normative values, are just plain wrong. 

\section{Free Will, Measurement Independence and Quantum Physics}
\label{quantum}

In this final Section we return to the question of whether (\ref{MI}) is equivalent to an assumption of free choice. With reference to the author's RaQM \cite{Palmer:2025a} It is claimed not. 

\begin{figure}
\centering
\includegraphics[scale=0.5]{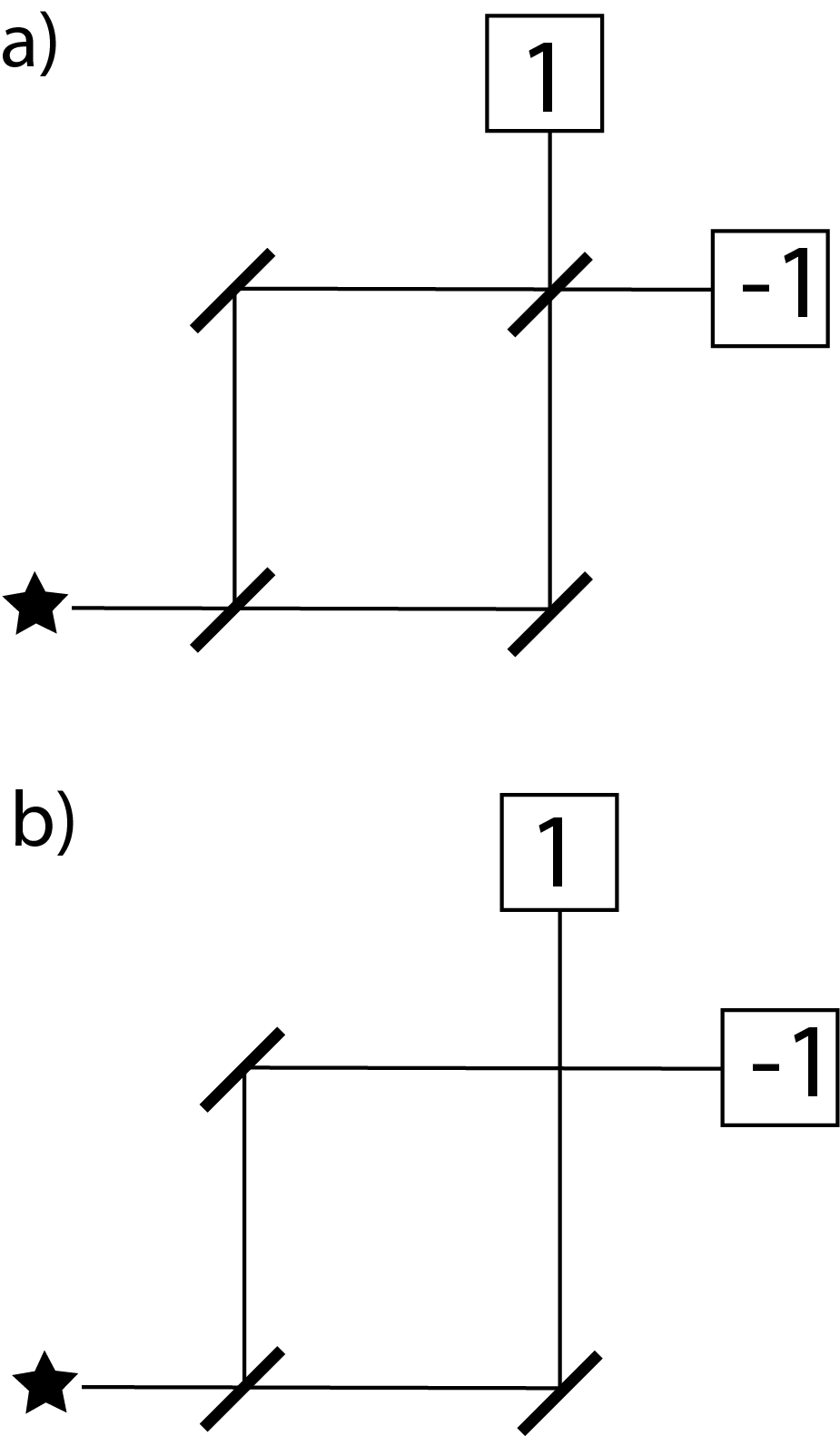}
\caption{\emph{a) A standard set up for the Mach-Zehnder interferometer, b) An experiment where the second half-silvered mirror has been removed. It is asserted that the experimenter is free to choose the set up, but having chosen, the counterfactual choice where the same particle is being measured, i.e., kept fixed, violates the rationality conditions in RaQM.}}
\label{MZ}
\end{figure}

Based on the discussion above, we have free will. Hence an experimenter is free to choose how to configure a quantum measurement. To be concrete, consider an Mach-Zehnder experiment (see Fig \ref{MZ}), in which the experimenter chooses between option $O=1$, where the second half-silvered mirror is in place leading to a wave-like interferometric experiment, or $O=0$ where the half-silvered mirror is removed leading to a particle-like `which-way' experiment. According to the model above, assuming $\alpha_n \ne 0 \mathrm{or} 1$, then in a situation where the experimenter chose $O=1$, they could have counterfactually chosen $O=0$, or vice versa.  

In quantum mechanics (QM), the wavefunction for a particle in the interferometer when $O=1$ can be written
\be
\label{O1}
|\psi_{O=1})\rangle = \cos \frac{\phi}{2}  |0\rangle + \cos \frac{\phi}{2}  |0\rangle 
\ee 
where $0<\phi<\pi/2 $ denotes a relative phase angle between the two arms of the interferometer. By contrast, when $O=0$
\be
\label{O0}
|\psi_{O=0}\rangle = \frac{1}{\sqrt 2}(|0\rangle + e^{i \phi} |1\rangle)
\ee
A key difference between QM and RaQM is that in RaQM, squared amplitudes and complex phases (in degrees) are required to be rational dyadic fractions. Hence in (\ref{O0}) we require $\cos \phi$ to be rational, whilst in (\ref{O1}) we require $\phi$ (in degrees) to be rational. By Niven's Theorem \cite{Niven} these two requirements are typically mutually contradictory: if $\cos \phi$ is rational, then $\phi$ is irrational, and vice versa. 

How do we interpret this incommensurateness in RaQM? Clearly an experimenter has some control over the relative length of two arms of the interferometer, and hence of the phase difference $\phi$. However, this control is necessarily nominal. Consider an interferometric experiment ($O=1$)) comprising a set $\{j=1,2,3, \ldots J\}$ of individual runs, each with a distinct particle. Then, although the values of $\cos \phi_j$ over each of the individual runs will agree to some nominal accuracy $(\cos \phi)^{\mathrm{nom}} \in \vmathbb Q$, the exact rational values of $\cos \phi_j -(\cos \phi)^{\mathrm{nom}}$ will differ from one run to the next. In \cite{Palmer:2025a} the exact value of $\cos \phi_j-(\cos \phi)^{\mathrm{nom}}$ provides a unique label for the $j$th particle. That is to say, the exact rational value $\cos \phi_j-(\cos \phi)^{\mathrm{nom}}$ can be considered a contextual hidden variable $\lambda_j$ for the $j$th particle associated with the $j$th run. 

This has profound implications. As discussed, for any given run, the experimenter is free to choose between an interferometric and a which-way experiment, and hence could have counterfactually chosen a which-way experiment if in reality they had chosen to perform an interferometric experiment. However, by the rationality conditions in RaQM, if $\cos \phi_j$ is rational in the real-world run, $\phi_j$ must be rational in the counterfactual-world run. But this means that $\phi_j$ cannot have the same value in the real-world run as the counterfactual-world run. But if the exact measurement setting provides a unique label, i.e. hidden variable, for the run, then according to RaQM it is not possible to have counterfactually chosen a which-way experiment, \emph{keeping fixed the particle that was measured in the real-world interferometric experiment}, keeping $\lambda_j$ fixed. We can restate this as follows. If an interferometric experiment took place in reality, then, by construction,
\be
\rho(\lambda_j | O=1) \ne 0
\ee
where $\rho$ is a probability density on the hidden-variables. According to RaQM, $\lambda_j=\cos \phi_j-(\cos \phi)^{\mathrm{nom}}$ is rational. As discussed, the experimenter was free to have performed a which-way experiment. However, necessarily in this counterfactual world, the phase $\phi_j$ must be a rational angle. Hence, in the counterfactual world
\be
\rho(\lambda'_j | O=0) \ne 0
\ee
where $\lambda'_j = (\phi_j-\phi^{\mathrm{nom}})/\pi$ is rational. However, by Niven's Theorem, it must be the case
\be
\rho(\lambda_j |O=0)=0
\ee 
That is, the Measurement Independence assumption
\be
\rho(\lambda_j |O=1)=\rho(\lambda_j |O=0)=\rho(\lambda_j
\ee
is false. In the counterfactual world where the experimenter chose the which-way option, the particle's hidden variable cannot be the same as in the real world where the experimenter chose the interferometric option. This counterfactual world violates the RaQM laws of physics. 

As discussed in \cite{Palmer:2025a}, \emph{exactly} the same arguments apply to Bell's Theorem, where now Alice chooses between the two measurement options $O_A=0$ and $O_A=1$, and Bob chooses between the two measurement options $O_B=0$ and $O_B=1$. As before, we allow Alice and Bob to each freely choose between their two options. However, having chosen,  the counterfactual worlds where either Alice and Bob individually choose otherwise, keeping the particles hidden variables fixed, is inconsistent with Niven's Theorem. In this way, for example
\be
\rho(\lambda |O_A=0\;O_B=0) \ne 0 \implies \rho(\lambda |O_A=0\;O_B=1)=0 \ \mathrm{and} \ \rho(\lambda |O_A=1\;O_B=0)=0
\ee
violating (\ref{MI}) but, importantly, not violating free choice. As discussed in \cite{Palmer:2025a} this leads to an interpretation of the violation of Bell inequalities in a deterministic model which is not EPR/Bell nonlocal, and yet where experimenters have free choice. 

One last point. According to the free-will theorem of Conway and Kochen \cite{ConwayKochen} not only do experimenters have free will so do the particles they measure. What does this mean? We have argued that just-in-time initialisation implies that choices are not simply functions of information from times earlier than the relevant just-in-time initialisation time, and hence not simply functions of the arbitrary past. If an experimenter chooses to make an interferometric measurement (so that $\cos \phi_j$ is rational), then there are clearly many possible exact values of $\cos \phi_j$ consistent with the rationality constraint and the nominal value $\phi^{mathrm{nom}}$ under the control of the experimenter. These exact values will be determined by information $Y_n$ in the just-in-time initialisations. Since these $Y_n$ are free variables, in the sense of Bell's quote in the Introduction, the particle hidden variables can be said to be free variables. 

\section{Conclusions}

While many of us have a very strong intuitive feeling that we could have done otherwise, i.e., have some sense of free will, the concept of free will is an especially vexing one for a scientist since it is hard to square it with either deterministic or indeterministic laws of physics. If the laws of physics are deterministic then everything we do is predetermined by initial conditions at the time of the Big Bang. If everything we do is associated with randomness and indeterminism, how can we `will' our decisions and actions. 

A deterministic model of free will (and decision making and moral responsibility) has been proposed which overcomes these objections. Based on the concepts of just-in-time initialisation for nonlinear systems with an upscale cascade of information from the Planck scale, and a stochastic-rounding with control strategy for decision making in the light of such information, we suggest that humans have genuine free will. Importantly, this allows us to understand a number of disparate aspects of the world and our place in it, from the existence of moral responsibility to understanding the experimental violation of Bell's inequality without violating the free choice assumption and without the requirement for EPR/Bell nonlocality. 

\section*{Appendix}

At first glance, it might seem impossible to describe the chain (\ref{Z2}) by a single timestep. For example, the propagation of information from sub-Planck to super-Planck scales presumably operates on the Planck timescale $10^{-41}$s. By contrast, the inherent timescale for a quantum object of mass $m$ is given by its Compton wavelength divided by $c$. For an electron the timescale $\approx 10^{-20}$s. By contrast, one might suppose a minimal time needed to physically implement some decision is around 1s. Each timescale differs from its predecessor by a factor of $10^20$. However, we can describe  (\ref{Z2}) by a single timestep by incorporating the effects of relativistic time dilation. If we associate the processes associated with a) sub-Planck to quantum, b) quantum to decision and c) decision to action with de Broglie waves
\be
e^{i(\omega_a t- k_a x)}; \ \ \ e^{i(\omega_b t- k_b x)}; \ \ \ e^{i(\omega_c t- k_c x)}
\ee
then we can consider three frames $\mathscr F_a$, $\mathscr F_b$ and $\mathscr F_c$ where the corresponding de Broglie wave is a standing wave $Ae^{i \omega_0 t}$ with common $\omega_0$. Then
\be
\omega_a =\gamma_a \omega_0; \ \ \ \omega_b =\gamma_b \omega_0; \ \ \ \omega_c =\gamma_c \omega_0
\ee
where $\gamma_{a}$ denotes the time-dilation factor associated with process a) etc. If we assume that $\gamma_c \approx 1$, then $\gamma_{b} \approx 10^{20}$, $\gamma_{b} \approx 10^{40}$.

\bibliography{mybibliography}
\end{document}